# Universal Quenching of Superconductivity in Two-dimensional Nano-islands

Jungdae Kim, Gregory A. Fiete, Hyoungdo Nam, A. H. MacDonald, and Chih-Kang Shih*

Department of Physics, The University of Texas at Austin, Austin, Texas 78712, USA

*Corresponding Author: shih@physics.utexas.edu

## Abstract

We systematically address superconductivity of Pb nano-islands with different thicknesses and lateral sizes via a scanning tunneling microscopy/spectroscopy. Reduction of the superconducting gap (Δ) is observed even when the island is larger than the bulk coherence length (ξ) and becomes very fast below ~ 50 nm lateral size. The suppression of Δ with size depends to a good approximation only on the volume of the island and is independent of its shape. Theoretical analysis indicates that the universal quenching behavior is primarily manifested by the mean number of electronic orbitals within the pairing energy window.

The remarkable properties of a superconductor are due to its Cooper pair condensate (formed from a macroscopic number of electrons), which can be described by a single quantum wave function. According to the celebrated Bardeen-Cooper-Schrieffer (BCS) theory of superconductivity[1], there is a minimum length scale (the coherence length, $\xi$) on which the condensate order parameter can vary. The fate of superconductivity in systems with spatial dimensions smaller than the coherence length, $\xi$, has been the subject of intense interest for decades because it is dependent on quantum confinement, interaction, and quantum coherence effects in an intrinsically many-particle context--ingredients that drive much modern research in quantum many-body physics[2-21]. Attention has often focused on two-dimensional (2-d) systems which can have fragile order because of quantum and thermal fluctuations. In this context, the recent discovery of *robust* superconductivity in epitaxial thin Pb films with thicknesses that are orders of magnitude smaller than $\xi$ seems surprising[4, 10, 11]. Even at a thickness of only two atomic layers (~ 0.6 nm), the superconducting transition temperature ($T_C$ ~ 5 K) of Pb films is only slightly smaller than the bulk value 7.2 K[4]. Nevertheless, there remain disputes regarding the thickness dependence of $T_C$ at thickness below 10 MLs. Although this observation is already interesting, much more telling information on the fate of superconductivity at small length scales emerges when the lateral dimensions of the ultra-thin films are also reduced.

What happens if all dimensions are reduced? Ultimately, at some length scale superconductivity should cease to exist. What determines this scale? Here we systematically address this fundamental question via a detailed scanning tunneling microscopy/spectroscopy (STM/STS) study of thin-film superconducting islands with different thicknesses and lateral sizes. By controlling the lateral size of ultrathin 2-d islands, we discovered that as the lateral dimension is reduced, suppression of the superconducting order occurs. The reduction of the superconducting order parameter starts slowly, but then accelerates dramatically when the level spacing starts to approach the value of the gap. While the values of the superconducting gap of the nano-islands show thickness dependence at all lateral sizes, when normalized to the extended film limit, they collapse to a universal curve that depends only on the volume of the nano-island. Most surprisingly, our results for the size of the gap Δ versus island volume show quantitative agreement with recent STS studies of hemisphere-shaped Pb droplets, despite their dramatic difference in geometry[5, 9].

The experiments were conducted in a home-built low temperature STM system with an *in-situ* sample preparation chamber. The striped incommensurate (SIC) phase of the Pb-Si surface shown in Fig. 1(a) was prepared by deposition of ~ 1ML of Pb onto the Si(111) $7\times7$ surface at room temperature, followed by sample annealing at 400 ~ 450 $^{o}$C for 4 min to form the surface template[22-24]. Depending on the detailed kinetic control, either flat film or 2-d nano-islands of Pb can be obtained. Flat films can be grown by holding the sample temperature at ~ 100 K during Pb deposition[4]. Fig. 1(b) shows a 5 monolayer (ML) film grown on the Pb-on-Si (111) SIC surface. In order to get 2-d islands with a variety of lateral sizes and thickness, Pb was deposited on the template at ~ 200 K with a deposition rate of 0.5ML per minute. Fig. 1(c) shows 2-d islands of different sizes and shape on the same substrate. Also shown in Fig. 1(d) is a close-up view of two 3ML 2-d islands, one with an effective diameter ($d_{eff}$) of 74 nm and one with an effective diameter of 15 nm. The effective diameter ($d_{eff}$) of each island is calculated by using $d_{eff} = \sqrt{ab}$, where *a* and *b* are lengths along the major and minor axes, respectively.

Fig. 2(a) shows *dI/dV* tunneling spectra acquired at 4.3 K for a 5ML film and 5ML 2-d islands of various effective diameters. Interestingly, even at a diameter of 235 nm, which is much larger than the bulk coherence length ξ (~ 80 nm), one already observes a small reduction of the superconducting gap in comparison to that of the extended film. As one can observe directly from the data, the trend toward gap reduction continues as the diameter of the 2-d islands decreases further. Fig. 2(b) shows *dI/dV* tunneling spectra acquired at 4.3 K for 3ML 2-d islands with diameters ranging from 15 nm to 74 nm. The size dependence of the superconducting gap of 4ML 2-d islands follows a similar trend to that of 3ML islands. It is important to recognize that, as long as the individual island is isolated, the measured superconducting gap is very uniform throughout the island until its very edge.

Recently Brun *et al.*[15] reported a study of thickness dependence of the superconducting gap of Pb 2-d islands with lateral size larger than ξ which they interpreted to represent the superconducting properties of extended films. They found that the superconducting gap scales with the inverse of island thickness and superconductivity should vanish at a thickness of 2 ML. While the recent observation of strong superconductivity at 2ML already implied the breakdown of the 1/d scaling[4], the current result that gap reduction occurs in 2-d islands with lateral dimension much less than ξ should settle the true behavior superconductivity in these nano-systems.

In order to more quantitatively characterize the suppression of superconductivity, we measured the transition temperature, $T_C$, for different lateral sizes and thicknesses of 2-d islands. To extract an energy gap value for different temperatures, $\Delta(T)$, we fit the normalized *dI/dV* spectrum measured by STS to BCS-like density of states as shown in Fig. 2(c). All STM signals were carefully shielded against radio frequency (RF) noise, and a Gaussian broadening parameter was implemented in the fitting to model the effect of the remaining noise. A standard deviation of 0.3 $mV_{rms}$ was used for the Gaussian broadening (see Appendix A for more details). To determine $T_C$, extracted values of $\Delta(T)$ are fitted to BCS gap equation (see fig. 2(d));

$$\frac{1}{N(0)V} = \int_0^{E_D} \frac{1}{\sqrt{\varepsilon^2 + \Delta(T)^2}} \tanh\left(\frac{\sqrt{\varepsilon^2 + \Delta(T)^2}}{2k_B T}\right) d\varepsilon \qquad (1)$$

where *N(0)* is the density of states at the Fermi level, *V* is the electron-phonon coupling, and $E_D$ is the Debye energy. Strictly speaking, the universal curve in $\Delta$ vs T is valid for a weak-coupling superconductor ( $E_D / k_B T_C >> 1$), but it is a good approximation in most cases, including Pb[5, 25].

Fig. 3(a) demonstrates that $T_C$ of 3ML, 4ML, and 5ML islands depends on their lateral size $d_{eff}$. Interestingly, there is a transition region (somewhere between 40 to 60 nm) below which electrons rapidly lose the strength of superconducting coherence. We also observe a variation of $T_C$ as a function of island thickness for a given $d_{eff}$ : $T_C$ (3ML) > $T_C$ (4ML) > $T_C$ (5ML), due to the quantum oscillations of $T_C$ from the vertical electronic confinement, a topic that has been intensively investigated recently[4, 19] (3ML data was previously unavailable due to difficulty in preparing 3ML films). The transition region from slow to rapid $T_C$ reduction also shows thickness dependence: it occurs at ~ 40 nm for 5ML islands, at ~ 50 nm for 4ML islands, and slightly above 50 nm for 3ML islands. On the other hand, if we normalize the $T_C$ of 2-d islands to the thin film value at a given thickness and plot it as a function of island volume, the data collapse onto a single curve, revealing a universal behavior of superconductivity suppression with respect to the island volume (Fig. 3(b)). Remarkably, this universal curve is in quantitative agreement with earlier STS studies for Pb droplets, despite the dramatic difference in geometry[5, 9]. This collapse demonstrates that superconductivity in regularly-shaped nano-islands is sensitive to a large degree only to the average level spacing of a nano-structure, affirming to a surprisingly degree theories[7, 8, 26] of finite-size superconductivity suppression in which the mean number of electronic orbitals within the pairing energy window plays the central role.

The data collapse of Fig.3(b) can be understood within BCS theory in the following way. Starting with the mean-field gap equation *at zero temperature*,

$$\frac{1}{V} = \sum_{n}^{\varepsilon_n = E_D} \frac{1}{\sqrt{\Delta^2 + \varepsilon_n^{\ 2}}} \, , \qquad (2)$$

where $V$ is the electron-phonon coupling as before, $\varepsilon_n$ are the discrete single-particle energy levels on the island, and $E_D$ is the Debye (phonon) energy as before, we can determine the dependence of $\Delta$ on the size of the island, which sets the spacing between the energy levels, $\varepsilon_{n.}$ Since the lateral size of the islands is finite, one should in general use the gap equation with a discrete sum, Eq.(2), rather than an integral, Eq.(1). However, if the islands are large enough, and one is looking at finite temperatures, the integral form (1) is a good approximation. That is the reason we have used it to fit our data earlier to obtain $\Delta$(T). Here, we wish to understand the fundamental reason that the suppression of superconductivity with lateral size follows a universal curve. For that purpose, we begin with zero temperature considerations and then generalize to finite temperature to obtain the universal curve theoretically (see the black curve in Fig.3(b)). To understand weakened superconductivity in small islands, one must retain the discrete nature of the energy levels as their increased spacing with shrinking lateral size plays an important role in the suppression of superconductivity[7, 8, 26]. We first note that our data show an absence of Coulomb blockade effects on the islands, which would have a charging scale of $E_C \sim e^2/C$, which can be estimated as $E_C$ ~10-40 meV. Here $e$ is the charge of the electron and the island capacitance $C=2\pi\,\kappa L$, where L is the size of the dot, and is $\kappa$ the dielectric constant. This indicates some electrical contact with the substrate (also needed for the STM measurement itself), though we are not able to determine if the coupling is in the intermediate or strong regime. Because the contact of the SIC is only a "ring" around the Pb island (i.e. the Pb island is not "sitting" on the SIC), even a highly conducting SIC-Pb interface would only weakly affect the basic superconducting properties which are primarily determined by the "bulk" island density of states and electron-phonon coupling. Thus, the essential physics of the universal suppression in our experiments is contained in Eq.(2), as we now show.

The typical scale of the variation in the density of states (number of energy levels on the island per unit energy) in Eq.(2) is set by the energy scale of the electronic degrees of freedom, $E_F$ (Fermi energy), which is 9.5 eV for Pb. On the other hand, $E_D$ is typically tens or hundreds of Kelvin (88 K in Pb), or roughly $10^{-2}$ electron volts. Because $E_D << E_F$, the density of states (*i.e.* average level spacing) will be constant to a good approximation over the energy range of the sum in (2), even for islands that are somewhat irregularly shaped. First consider an island with a shape that supports perfectly evenly space levels. (We will later show that when the energy levels are not perfectly spaced, there is very little change in the physics expressed through a self-consistency condition in the BCS gap equation, Eq.(2)—this is ultimately what is responsible for the universal suppression of the superconductivity.) When fluctuations in level separations are neglected, $\varepsilon_n = n\,\delta$, where $\delta$ is the level spacing. The corresponding density of states for such an island is thus $1/\delta$.

From the well-known formula for the bulk gap in the thermodynamic limit, $\Delta_0=2E_D\exp\{-1/\lambda\}=2E_D\exp\{-\delta/V\}$, and the bulk value of the transition temperature, $T_C=\Delta_0/2.2=7.2$ K, we find $\delta/V=2.5$ for Pb under the equal level spacing assumption $\varepsilon_n=\delta\, n$. Here V is the strength of the electron-phonon coupling in units of energy. Thus, for the critical island size where the gap $\Delta = 0$, we have

$$2.5 = \frac{\delta}{V} = \sum_{n=1}^{n_{\max,0}=E_D/\delta} \frac{1}{n}\,, \tag{3}$$

which implies that $n_{max,0}=7$ when the island becomes so small that the transition temperature vanishes, or that $\delta = E_D/7 \sim 13$ K is roughly of order the bulk gap, $\Delta_0 = 12$ K. The key analytical feature of Eq.(3) is the "1/n" contribution from the different energy levels: Even for "large" n, there are appreciable contributions to the sum (1/n is logarithmically divergent). This means that any random (not too large) fluctuations from the equal-level spacing approximation due to an irregular shaped island will be approximately averaged out. (We have verified this in numerical checks with random energy spacings constrained to have the same average.) Moreover, in islands larger than the critical size (with smaller level spacing $\delta$) there will be *even more effective averaging* because the larger "small" n contributions will carry less weight due to the presence of the gap in the denominator of Eq.(2) and Eq.(4), and there will be more terms in the sum to be averaged, $n_{max} > n_{max,0}$:

$$2.5 = \frac{\delta}{V} = \sum_{n=1}^{n_{\max}=E_D/\delta} \frac{1}{\sqrt{\left(\frac{\Delta}{\delta}\right)^2 + n^2}} \tag{4}$$

From Eq.(4), the gap dependence on level spacing, $\Delta(\delta)$, is determined. We emphasize this analysis is valid even in the presence of tunneling coupling of the island states to the substrate, with only a small numerical change in the results. The arguments above are not simply hand waving, but rather rely on the properties of the summation of 1/n for a finite number of terms.

As we now show, similar considerations apply to the level spacing dependence of the *critical temperature*, $T_C$, which is proportional to the gap. Formally, the temperature dependence of the gap in a small island is given by the discrete version of Eq.(1),

$$2.5 = \frac{\delta}{V} = \sum_{n=1}^{n_{\max}=E_D/\delta} \frac{Tanh[\sqrt{(\Delta(T))^2 + (n\delta)^2}/(2k_B T)]}{\sqrt{\left(\frac{\Delta(T)}{\delta}\right)^2 + n^2}} \tag{5}$$

which will exhibit the same insensitivity to level spacing fluctuation as Eq.(4). The critical temperature, $T_C$ , is given by Eq.(5) with $\Delta(T_C)=0$,

$$2.5 = \frac{\delta}{V} = \sum_{n=1}^{n_{\max}=E_D/\delta} \frac{Tanh[n\delta/(2k_B T_C)]}{n} \tag{6}$$

For islands of equal volume, but different shape, the value of $n_{max}$ in Eqs.(6) will be roughly the same, since that only counts the total number of states in the energy window within $E_D$ of the Fermi energy, and that number is expected to be a weak function of island shape (distorting a regular shaped island into an irregular shape mainly rearranges energy levels, which is unimportant because of the "averaging" effect from the slow decay of the summand in Eqs.(5)). We believe the effective averaging of level spacing fluctuations due to a roughly constant *average* density of states over the relevant energy range, $E_D$, is the reason we find that the normalized critical temperature curves in Fig.3(b) collapse to a universal form that is to a good approximation independent of island shape and given by Eq.(6). We remark that first principles calculations have predicted some changes in the phonon spectrum and electron-phonon coupling, but this makes a quantitatively small change in the transition temperature in the thin-film geometry[19]. We believe our argument based on the BCS mean-field gap equation simply explains all the trends in our data and related experiments[5, 9].

In summary, we have performed a STM/STS study of 2-d superconducting islands with different thicknesses and lateral sizes. As the lateral dimension is reduced, the strength the of superconducting order parameter is also reduced, first slowly at a dimension larger than the bulk coherence length, then dramatically at a critical length scale of 40 ~ 50 nm, which corresponds to level spacing of order the bulk gap $\Delta$. Interestingly, the systematic suppression of superconductivity with size depends to a good approximation only on the volume of the island and is independent of its shape. We have explained this feature with theoretical arguments based on BCS theory. We expect this work to have broad implications for device implementation that depends on detailed knowledge of size dependent superconductivity, and to stimulate further fundamental studies on nanoscale superconductivity.

## Acknowledgements

This work was supported by NSF Grant No. DMR-0906025, CMMI-0928664, Welch Foundation F-1672, and Texas Advanced Research Program 003658-0037-2007. GAF acknowledges support by ARO W911NF-09-1-0527 and NSF DMR-0955778.

## Appendix A: Determination of the superconducting gap

### 1. Normalization of tunneling spectra

Due to the existence of quantum well states, the density of states (DOS) of the sample is often not a constant within the relevant energy window of $E_F \pm 20$ meV. Consequently, the raw *dI/dV* data contains an asymmetric background. Moreover, as will be discussed in appendix II, there exist "pseudogap" features (the depression in *dI/dV* in the range between -10 mV to 10 mV) at temperature above $T_C$[16]. These two factors--the non-constant DOS near $E_F$ and the pseudogap--need to be considered in the normalization procedure. In most cases, one can normalize the spectra by dividing them with the spectra acquired above $T_C$ which should represent the normal state DOS. However, due to the existence of the pseudogap such a normalization procedure will artificially raise the *dI/dV* value in the vicinity of the superconducting gap, thus distorting the lineshape (see Fig. 4(a,b)). We found that the spectra normalized in this manner would not fit well with a BCS-like gap function (discussed below). Another possibility is to use spectra acquired at a temperature much above $T_C$ (say, > 80K) to represent the normal state DOS. However, the thermal drift encountered with such a large temperature change makes it difficult to guarantee that the tip will stay at the same location with respect to the sample.

Here we use a modified normalization procedure by using a high order (in this case $5^{th}$ order) polynomial to fit the spectra outside the gap region to represent the normal DOS (see the dashed line in Fig. 4(a)). The resulting normalized spectra are shown in the red line of Fig 4(b). This normalization procedure works quite well for a general shape even when the normal DOS contains a dip or peak. Moreover, it can be incorporated into a computer program that can automatically generate the normalized spectra without human bias. We recognize, however, the existence of pseudogap degrades the precision of fitting a small gap for spectra acquired at very close to Tc). Consequently, we regard those gap values below 0.2 meV as unreliable and are ignored in the $T_C$ fitting.

## 2. Gap fitting

While the BCS gap function[1] is not strictly applicable to the case of strong-coupling like Pb, it remains a good approximation. Within this approximation, the differential conductance *(dI/dV)* based on the tunneling current between a normal metal (STM tip) and a superconductor (sample) can be expressed as

$$\frac{dI}{dV} \propto \int_{\Delta}^{\infty} \frac{|E|}{\sqrt{E^2-\Delta^2}} \left[ -\frac{df}{dE}(E+eV) \right] dE \tag{A1}$$

where *f(E)* is the Fermi distribution function, *V* is an applied voltage bias, and Δ is the superconducting energy gap. In addition, a Gaussian broadening of finite width (called the broadening parameter) is applied. This broadening is due to the incomplete shielding of the radio frequency (RF) interference present at the tunneling junction as we discuss below.

For each RF shielding configuration, there is only one broadening parameter. This parameter changes when the configuration of the RF shielding changes. Shown in Fig. 5(a) and (b) are results of STS measurement of 9ML, without the RF filter and with our best filter configuration we have achieved so far, respectively. For the no-filter configuration, a 0.8 meV broadening parameter is needed to fit the spectra, corresponding to a RF noise with rms amplitude of 0.8 mV. On the other hand, with a good RF shielding configuration, only 0.2 meV broadening is needed. For a comparison, we also show the theoretical spectra at different temperatures without any broadening in Fig. 6(a). One can see that the spectra acquired with our best RF shielding configuration are reasonably close to the theoretical curves. It should be noted that in the temperature range where the spectra were acquired, the width of the Fermi-Dirac distribution already exceed the broadening parameter typically used in our experiments.

Most importantly, while the raw data of the two configurations differ significantly in their line shape, they yield a very similar $T_C$ (to within ± 0.1K). In Fig. 6(b), we the fit the gap values as a function of temperature from these two different configurations (shown as empty and solid squares). As one can see from this plot, the data from different filter configurations fall quite well onto a single BCS-like $\Delta(T)$ curve.

The best RF shielding configuration used in the experiment unfortunately compromises the operation of the (low-temperature) LT walker, which is an important component of LT-STM. Thus, for most of the experimental results reported in the main text, the spectra were acquired in a different RF filter configuration, corresponding to a 0.3 meV broadening parameter. This second RF filter configuration allows us to simultaneous operate the UHV LT walker. Since we have shown that the fitted gap and the resulting $T_C$ are independent of the broadening parameter being used to within ± 0.1K, we believe the scientific conclusions would be consistent. We should also emphasize that the temperature dependence of the gap near $T_C$ provides the most reliable determination of $T_C$ because the gap varies most rapidly close to $T_C$. This can be clearly seen by observing the raw data of a 5ML film in Fig. 6(c): while at 5.8 K, the superconducting gap is still clearly observable, it disappears at 6.3 K, consistent with the fitted $T_C$ value of 6.1K.

## Appendix B: The pseudogap structure at voltages above the gap

The majority of this paper is focused on the superconducting properties of nano-islands of different thicknesses and shapes. However, our data also show interesting features at voltage biases above the gap when the temperature is below $T_C$, and for all voltages we probed when the temperature is above $T_C$. As seen in Fig. 7 (a) and (b) (where the temperature is above $T_C$), there is a "pseudogap" feature that appears (even for very large islands). The pseudogap does not measurably change when the temperature drops below $T_C$ suggesting that it is likely an intrinsic property of the normal state of the Pb islands and not related to superconductivity in any way. It has been suggested in an earlier study[16] on films that such features may be due to phonon effects. We show here that we obtain a very good fit to a theory based on a pseudogap feature arising only from the combined influences of electron-electron interactions and disorder[27]. Moreover, we are able to extract a quantitative estimate of the *normal state* conductivity of the Pb nano-islands used in our experiments.

According to Ref 26, the tunneling density of states should follow a form

$$\frac{dI}{dV} = c_1 + c_2 |V|^{\alpha} + mV \tag{B1}$$

where $c_1$, $c_2$, $\alpha$ and m all depend on details of the system, and V is the voltage. The final term mV accounts for a weakly energy-dependent tunneling matrix element between the STM tip and the Pb substrate rather than intrinsic island properties and does not enter our estimates of island conductivity. For our purposes, we are most interested in $\alpha$ as it is inversely proportional to the 2-d conductance of the nano-island/thin film substrate:

$$\alpha = \frac{e^2}{h\sigma} \frac{\ln(2\pi a e^2 dn / d\mu)}{2\pi} \tag{B2}$$

where *e* is the charge of the electron, *h* is Planck's constant, $\sigma$ is the 2-d conductivity of the island or film, *a* is the tip-sample spacing, and $dn/d\mu$ is the compressibility of the island or film. (Here n is the 2-d density and $\mu$ the chemical potential.) Since the expression is only logarithmically dependent on the compressibility, we can estimate it by using a free electron approximation. For 2-d electrons, $dn/d\mu=4\pi m^*/h^2$, where $m^* = 1.14\ m_e$ is the effective mass of Pb and $m_e$ is the bare mass of the electron[28].

The plots of conductance vs island size are given in Fig. 7(c,d) for 3 ML and 4 ML film thicknesses. The values are consistent with those measured in early film samples[29, 30], *but here we provide the first conductance measurements of islands*. The general trend is for thicker and larger islands to be more conducting, and for the conductance to be higher as the temperature is lowered. These findings are all consistent with expectations. Fig. 8 shows that the pseudo-gap feature at energies above the gap does not change as the temperature drops below the superconducting transition temperature.

## Figure legends

### Figure 1

STM topography image of (a) the striped incommensurate (SIC) phase of Pb-Si surface, (b) a globally uniform 5ML Pb film, and (c) 2-d Pb islands with various lateral sizes and thicknesses (sample bias $V_s$ = (a) 0.3 V, (b,c) 2 V, tunneling current $I_t$ = (a) 20 pA, (b,c) 10 pA). The numbers labeled on the image (c) indicate thickness of islands in monolayer. (d) Two 3ML Pb islands with effective lateral size ($d_{eff}$) of 74 nm and 15 nm (inset) are shown in the same length scale (sample bias $V_s$ = 0.3 V, tunneling current $I_t$ = 10 pA).

### Figure 2

Lateral size dependence of differential conductance spectra (*dI/dV*) measured at 4.3 K is shown for (a) 5ML islands and (b) 3ML islands. All differential conductance spectra were taken with the same tunneling parameter with the junction stabilized at $V_s$ = 20 mV and $I_t$ = 30 pA tunneling current. (c) Normalized conductance spectra (black) measured as a function of temperature from a 3ML Pb island with a 62nm lateral size were fitted using the BCS density of states for the tunneling conductance (red). Each normalized spectrum is offset successively by 0.5 for clarity. (d) The superconducting energy gaps ($\Delta$) for each temperature were obtained from (c) and plotted as red squares. The blue curve is a fitting of these energy gap data using a BCS gap equation to obtain a $T_C$ of ~ 5.6 ± 0.1 K for a 3ML island with 62 nm lateral size. All $T_C$ values determined from such fitting are estimated to have an error bar of ± 0.1 K.

**Figure 3**

(a) $T_C$ as a function of island lateral size for 3ML, 4ML, and 5ML islands. A transition region from slow to dramatic reduction of superconductivity is shaded in green. (b) $T_C$ of islands was normalized to the globally flat film $T_C$ for each thickness and plotted as a function of cube root of island volume. For the $T_C$ normalization, 5ML film $T_C$ was measured to be 6.1 K, and estimated values of 6.9 K and 6.3 K from the $T_C$ trend in (a) were used for 3ML and 4ML film, respectively.

**Figure 4**

(a) *dI/dV* tunneling spectra obtained from 9ML Pb films at 4.3 K and 8.5 K. (b) The result of normalization with the spectrum acquired at 8.5 K (blue line) and the 5th order polynomial fitting (red line).

**Figure 5**

Temperature dependence of superconducting gap spectra taken from 9ML Pb films (a) without RF filter and (b) with RF filter. 0.8 meV and 0.2 meV broadening were applied in the BCS density of states (red lines) of (a) and (b), respectively. Each normalized spectrum is offset successively by 0.3 for clarity.

**Figure 6**

(a) Direct comparison between experimental spectra measured from 9ML Pb films with our best RF shielding configuration and theoretical BCS density of states without any broadening. Each spectrum is offset successively by 0.4 for clarity. (b) The superconducting energy gaps ($\Delta$) obtained from Fig. 5(a) and (b) were plotted as empty and solid blue squares for each temperature, respectively (error bar for each $\Delta$ value is smaller than the size of square mark). The red curve is a fitting of these energy gap data using a BCS gap equation. Those gap values from two shielding configurations align very well with a single BCS curve within a $T_C$ of ~ 6.6 ± 0.1 K. (c) Temperature dependent gap spectra measured from 5ML Pb films.

**Figure 7**

(a,b) Comparison of *dI/dV* spectrum measured at 6 K (above $T_C$) for 3ML islands and theoretical tunneling density of states based on electron-electron interaction and disorder (Eq. (B1)). Since T = 0 K is assumed in equation (B1), 1.2 meV broadening effect is applied to the fitting curves to take finite temperature and instrument noise into account. (c,d) 2-d conductivity ($\sigma$) values for Pb islands obtained from the fitting result in (a,b).

**Figure 8**

Shown is a comparison of the pseudo-gap feature at temperatures below and above the superconducting transition temperature. The lack of any clear temperature dependence for energies above the gap (clear from fit to Eq.(B1)) indicates the high-energy part of the pseudo-gap is not related to superconductivity, but is rather a normal state property of the Pb nano-islands.

# Figure 1

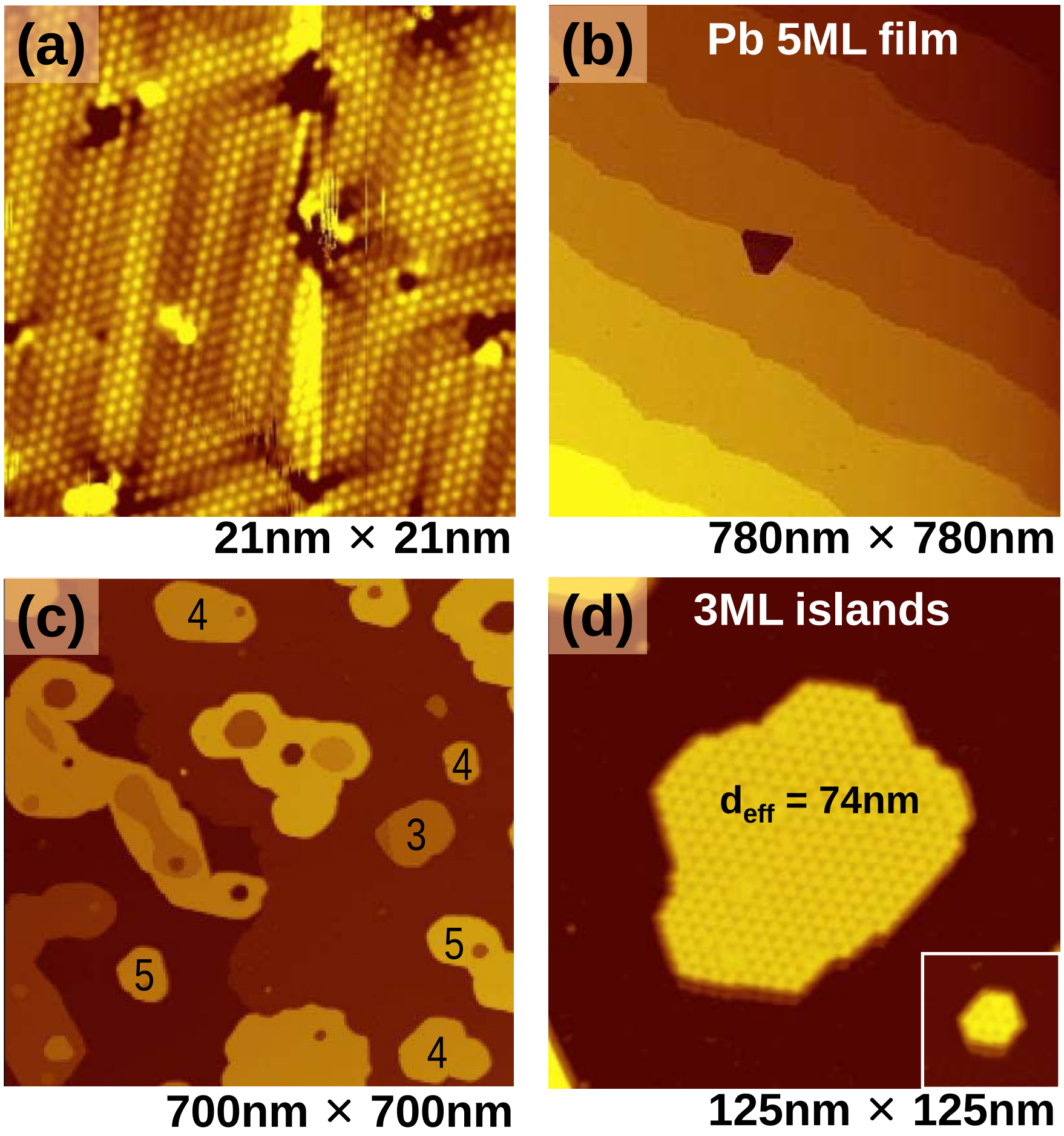

**Figure 2**

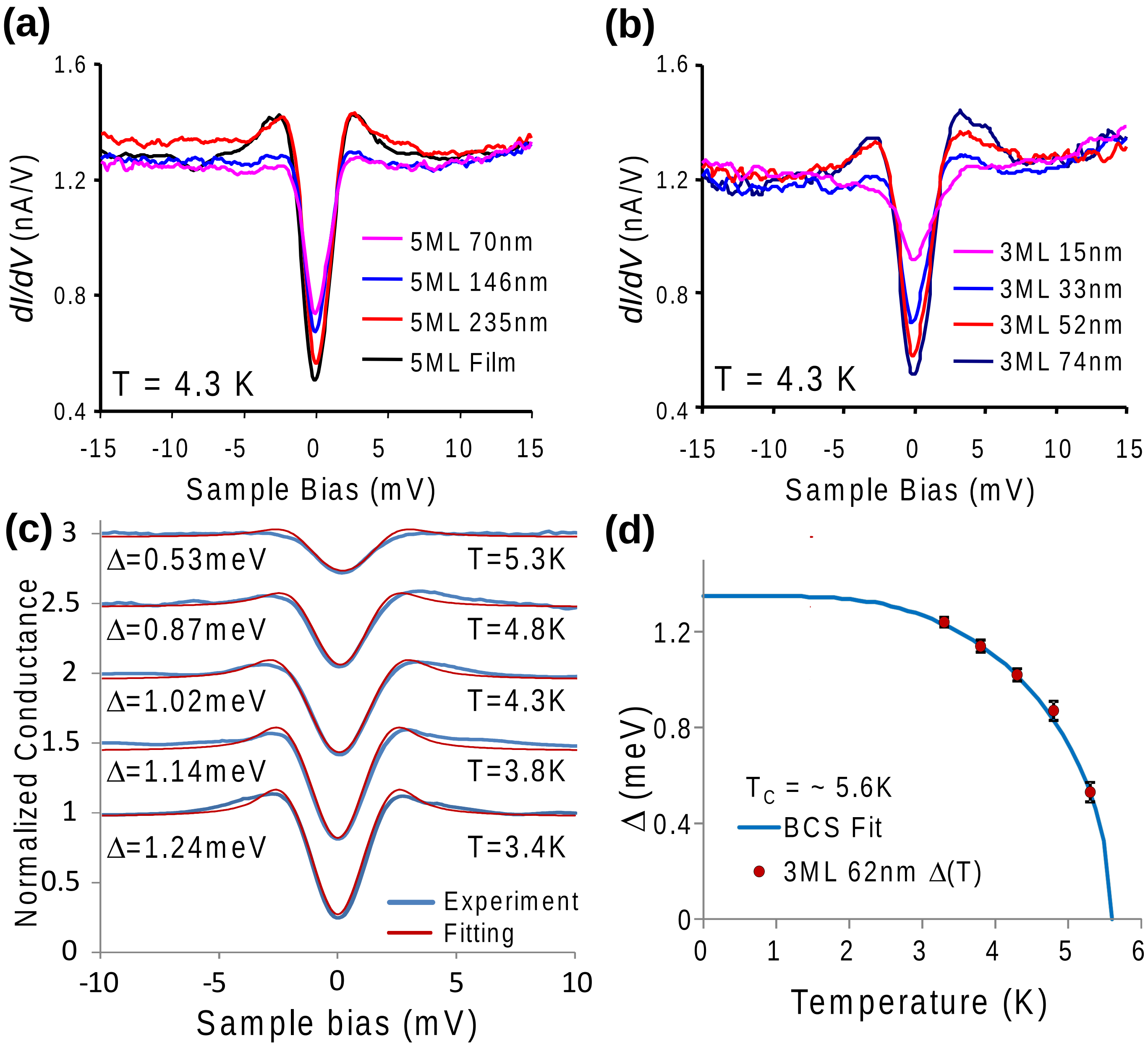

**Figure 3**

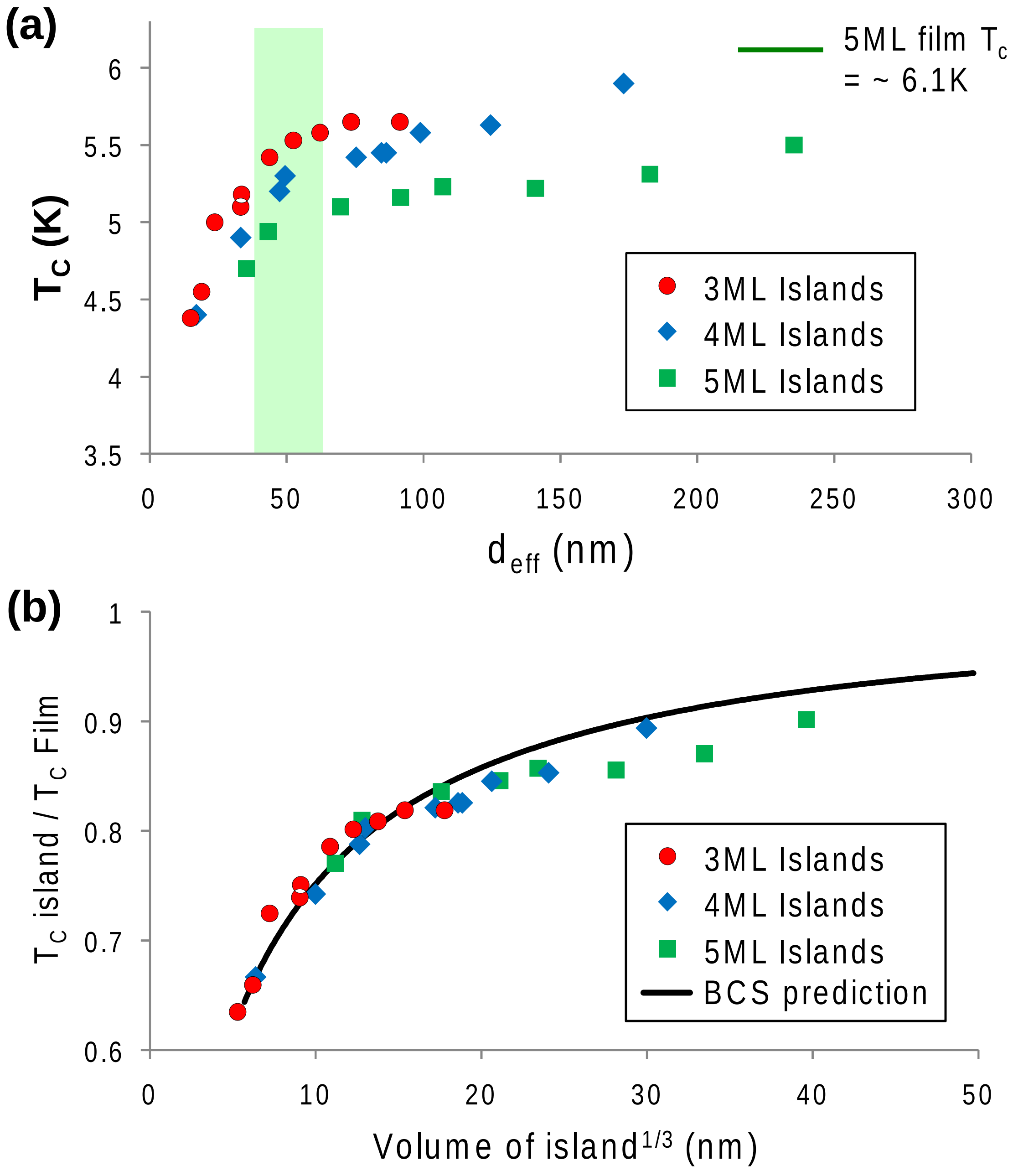

**Figure 4**

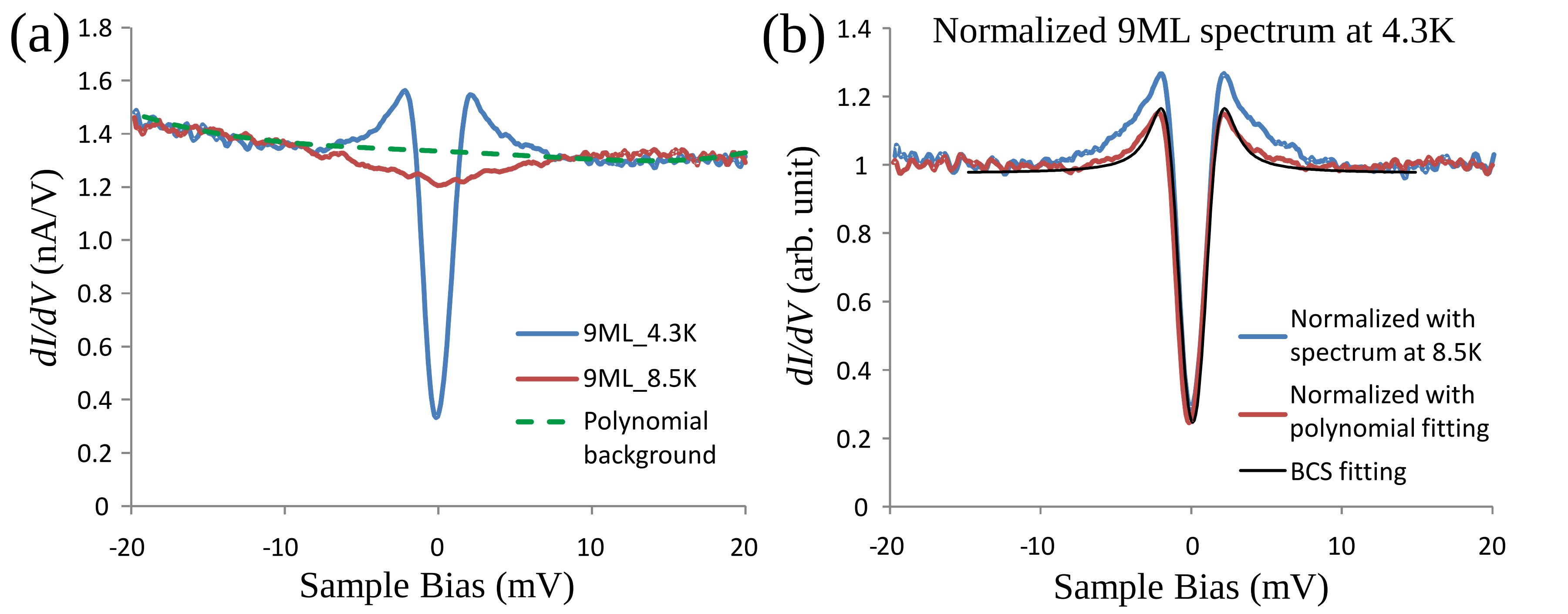

**Figure 5**

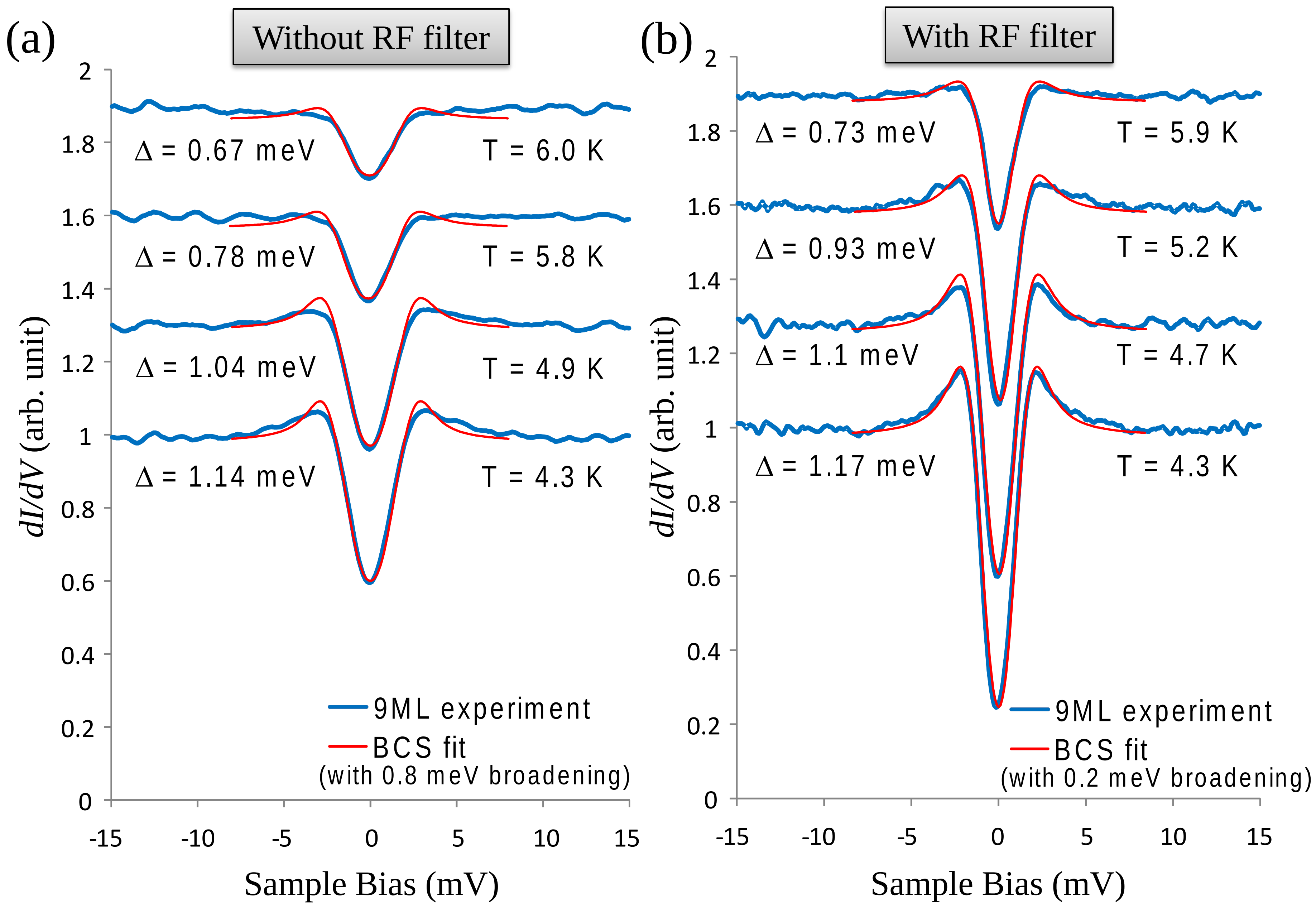

**Figure 6**

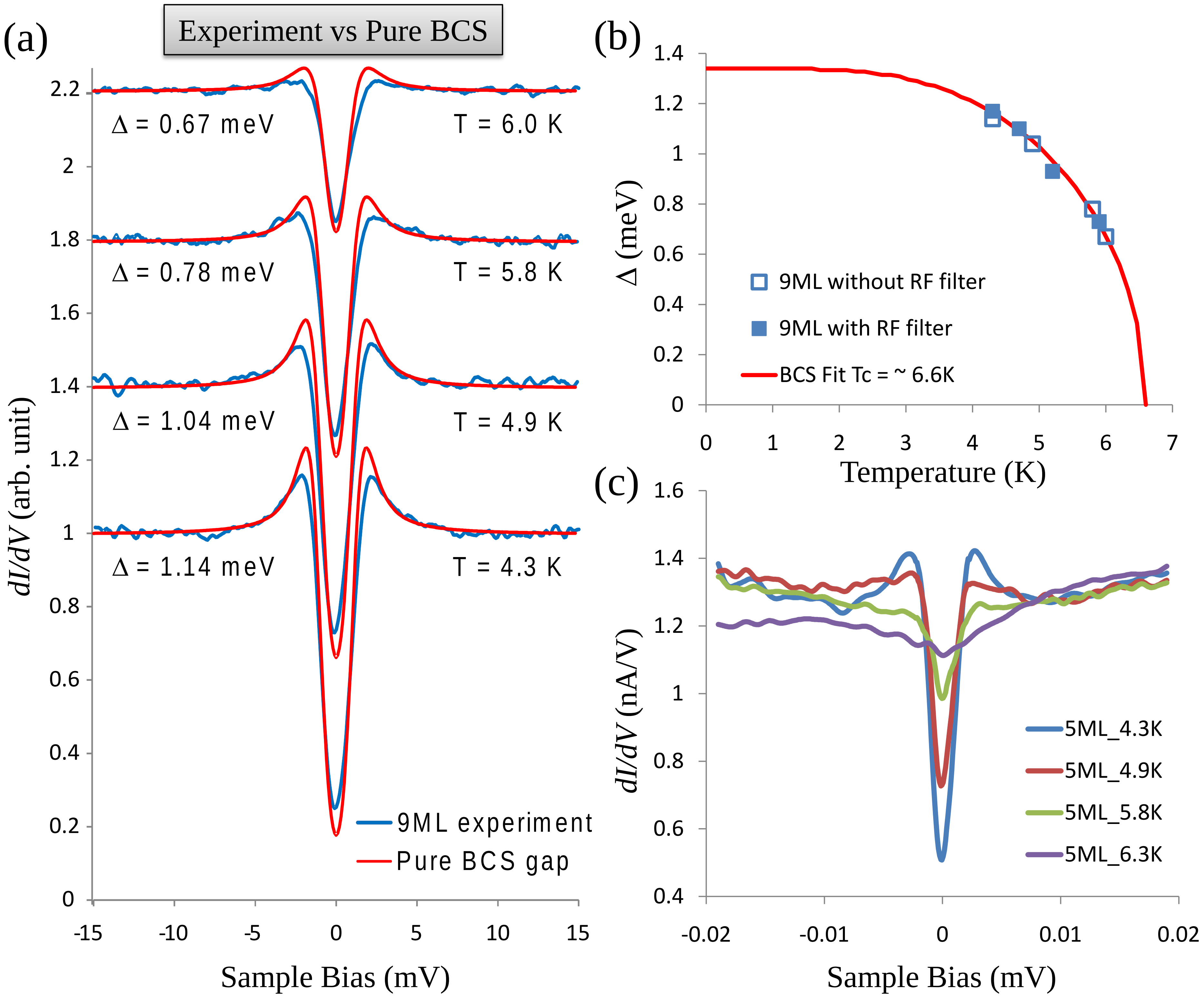

(a)
Experiment vs Pure BCS
Δ = 0.67 meV
T = 6.0 K
Δ = 0.78 meV
T = 5.8 K
Δ = 1.04 meV
T = 4.9 K
Δ = 1.14 meV
T = 4.3 K
9ML experiment
Pure BCS gap
dI/dV (arb. unit)
Sample Bias (mV)
(b)
9ML without RF filter
9ML with RF filter
BCS Fit Tc = ~ 6.6K
Δ (meV)
Temperature (K)
(c)
5ML_4.3K
5ML_4.9K
5ML_5.8K
5ML_6.3K
dI/dV (nA/V)
Sample Bias (mV)

**Figure 7**

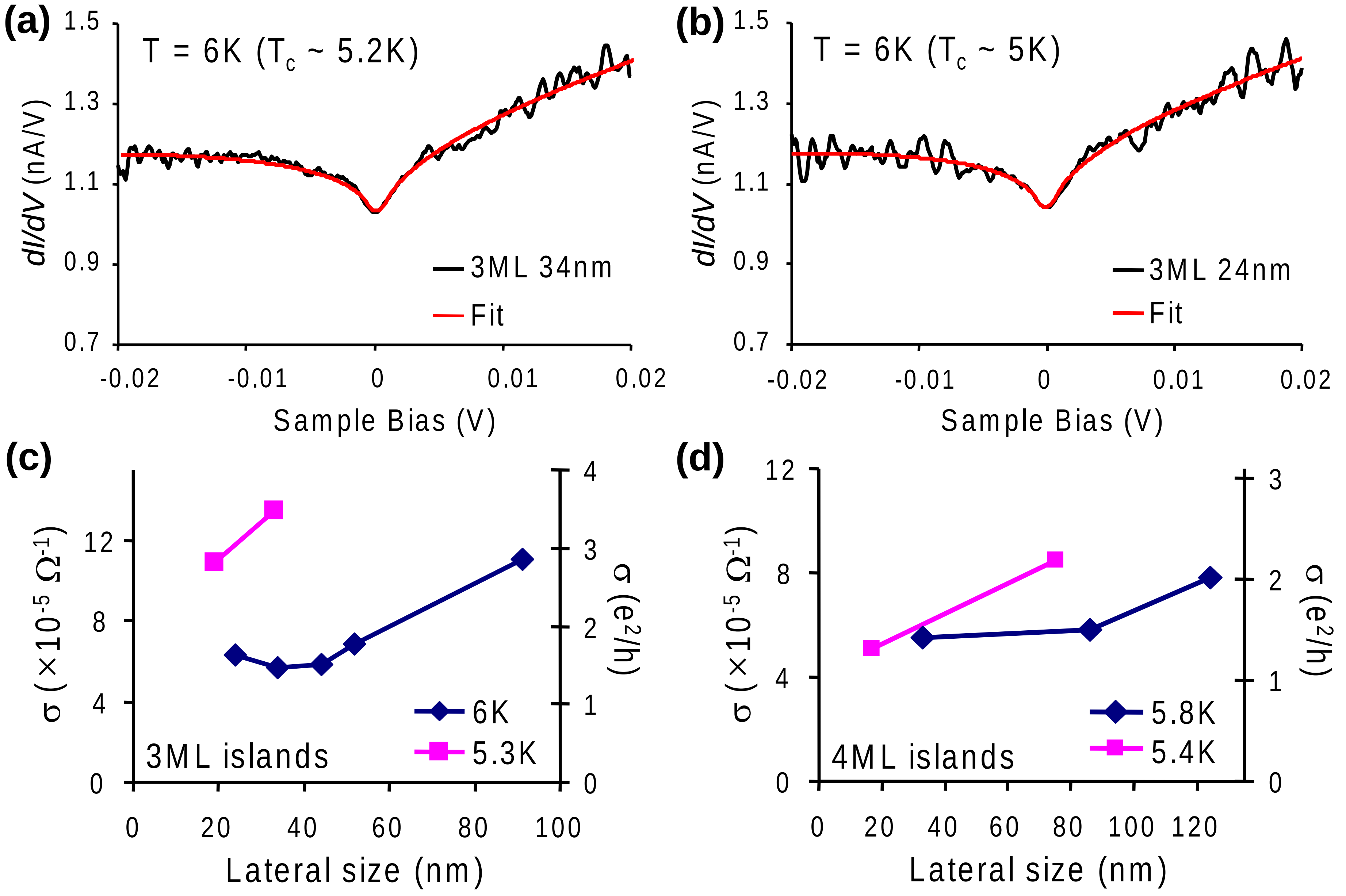

**Figure 8**

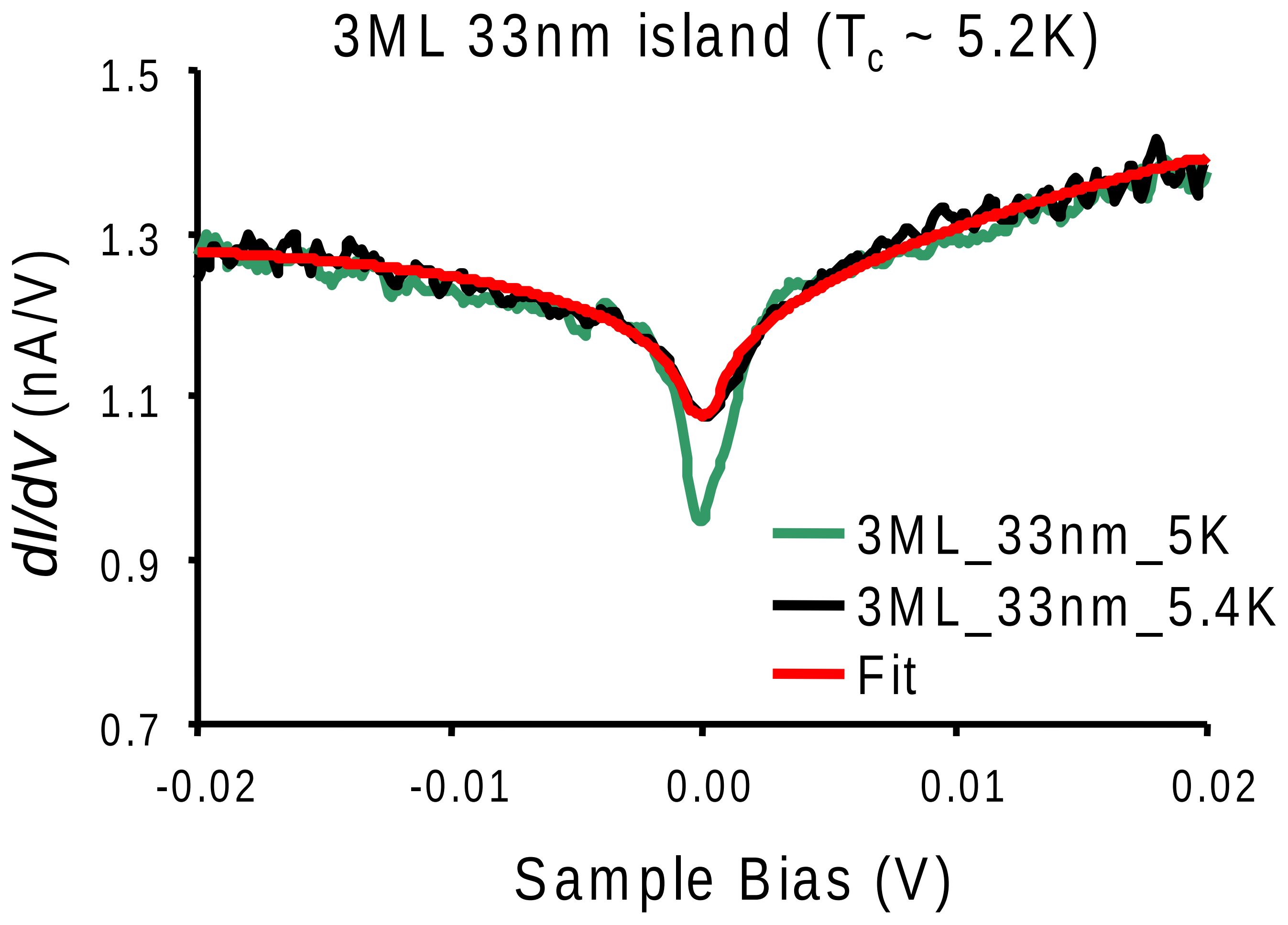